\documentclass[11pt]{article}
\usepackage{graphicx}
\usepackage{amssymb}
\usepackage{a4}
\begin{document}
\author{M. Bellon and M. Talon
\thanks{L.P.T.H.E. Universit\'es Paris VI--Paris VII (UMR 7589),
 Bo\^{\i}te 126, Tour 16, $1^{er}$ \'etage,
 4 place Jussieu, F-75252 PARIS CEDEX 05}
}
\title{Spectrum of the quantum Neumann model}
\date{June 2004}
\maketitle
\begin{abstract}
We study numerically the spectrum and eigenfunctions of the quantum
Neumann model, illustrating some general properties of 
a non trivial integrable model.
\end{abstract}
\vfill
LPTHE--04--12
\eject

\section{Introduction.}

The Neumann model, one of the first classical models to be shown
integrable in a non trivial manner~\cite{Neumann}, consists of a point
constrained to live on a sphere, and submitted to different harmonic
forces along the coordinate axis. It has been further studied, generalized
to higher dimension and related to the Jacobi theory of geodesic flows
on the ellipsoid by J. Moser~\cite{Moser}, while K. Uhlenbeck~\cite{Uhl}
found the conserved quantities in involution as required by Liouville
theory.  The classical Neumann model has been the object of much interest
since its spectral curve is the generic hyperelliptic Riemann surface.
Mumford~\cite{Mu84} has been able to provide nice derivations of classical
results on hyperelliptic curves by elaborating on this model and to give
explicit formulae for its solution.

Quantization of this model is non-trivial due to the constraints on the 
coordinates. This problem has been solved and the commutation of the natural 
quantization of the conserved quantities has been proved in~\cite{AvTa90}.
This result was not obvious, but can be understood naturally at present in the
framework presented in~\cite{BaTa03}, which asserts that under fairly general
conditions, classical integrable systems have integrable quantization.

The common spectrum of the conserved quantities has been studied
in~\cite{BaTa92}, where it was shown that the Schr\"odinger equation separates
using Neumann's coordinates. The problem reduces to the solution of a single
differential equation involving the common set of eigenvalues, and one has to
impose univaluedness of the wave function on the sphere to fix the
eigenvalues. This is a difficult problem since, although the differential
equation  looks like a simple generalization of the hypergeometric equation,
it cannot be solved by generalizations of Riemann's integral formulae, hence
one cannot compute the monodromies that would be necessary to state the
eigenvalue equations in closed form. However, writing the semi-classical
Bohr--Sommerfeld conditions is manageable, and has been achieved
in~\cite{BaTa92}. This line of approach has been developed further by D.~Gurarie
in~\cite{Gurarie}.  Introducing the Maslov indices for the classical
trajectories, he has been able to refine the semi-classical quantization
condition. He further considers the perturbative expansion around the trivial
case where all oscillator forces vanish, i.e., one deals with free motion on a
sphere.

However, all these works fell short of giving concrete solutions for the
spectrum of the quantum Neumann model. In fact, almost no non-trivial quantum 
integrable model yields exact formulae for the spectrum, the Calogero model
being a notable exception.  The aim of this paper
is to present a fairly detailed study of
the spectrum and eigenstates of the quantum Neumann model, obtained through
numerical methods. The problem has been made manageable by the recognition
that, for the model on $S^{N-1}$, $N$ parity operators commute with all
conserved quantities. This translates in a unique set of boundary conditions
for the factors in the separated wave function, so that these factors are
slices of a unique solution of the separated Schr\"odinger equation.
A local study allows to translate the numerical problem to a multiparameter
spectral problem with an assorted number of boundary conditions, for which the
COLNEW program yields a very efficient solution.

We shall first recall the definition of the classical Neumann model,
introducing a two by two Lax pair formulation. This allows to introduce in a
natural way Neumann's separating coordinates and Uhlenbeck's conserved
quantities. We then describe the formulation
of the quantum problem, followed by a presentation of the numerical methods. 
Finally, we show the first thirty-six energy levels of the system plotted
as a function of the strength of the potential.

\section{The Neumann model.}

We begin by presenting the classical Neumann model, which is discussed
in a different manner in the book~\cite{BBT}.
We start from a $2 N$-dimensional free phase space $\{ x_n, y_n,\quad
n = 1 \cdots N \}$ with canonical Poisson brackets: $\{x_n,y_m\} =
\delta_{nm}$ and introduce the ``angular momentum'' antisymmetric
matrix: $J_{kl} = x_k y_l - x_l y_k$ and the Hamiltonian:
\begin{equation}
H = {1 \over 4} \sum_{k \neq l} J_{kl}^2 ~+ ~ 
{1 \over 2} \sum_k a_k x_k^2
\label{hamiltonian}
\end{equation}
We shall assume in the following that: $a_1<a_2<\cdots<a_N$.
With the vectors $X =(x_k)$ and $Y = (y_k)$
and the diagonal constant matrix $A =(a_k \delta_{kl})$, the Hamiltonian
equations read:
$$\dot{X} = - J X \qquad \dot{Y} = -J Y - A X$$
They automatically ensure that $r^2=\sum x_k^2$ remains constant
and lead to the non-linear Newton equations for the particle:
$$\ddot{x}_k = - r^2 a_k x_k - x_k \sum_l ( \dot{x_l}^2/r^2 - a_l
x_l^2 )$$

The Liouville integrability of this system is a consequence of the
existence of $(N-1)$ independent quantities in involution~\cite{Uhl}:
\begin{equation}
F_k = x_k^2 + \sum_{l \neq k} {J_{kl}^2 \over a_k - a_l}
\label{conserved}
\end{equation}
This can be understood easily when one introduces the following Lax
pair formulation of the Neumann model: 
\begin{equation}
L(t)=L_0+\sum_k {1\over t - a_k} L_k,\quad L_0=\pmatrix{0&0\cr-1&0},\quad
L_k=g_k \pmatrix{0&1\cr 0&0}g_k^{-1}
\end{equation}
As usual in such a formulation, all dynamical variables are described
by the $g_k$. Since $L_k$ is nilpotent, $g_k$ can be taken as living
in the group of lower triangular matrices of unit determinant and we 
parametrize it as:
\begin{equation}
g_k=\pmatrix{x_k&0\cr y_k&x_k^{-1} }
\end{equation}
The matrix $L(t)$ can therefore be written:
\begin{equation}
L(t) = \pmatrix{-v(t)&u(t)\cr -w(t)&v(t)} 
\end{equation}
with
$$ u(t) = \sum_k {x_k^2 \over t - a_k},\ v(t) = \sum_k {x_k y_k\over t - a_k},\
w(t) = 1 + \sum_k {y_k^2 \over t - a_k} $$
The spectral curve, $\det (L(t) -s I)=0$, reads 
\begin{equation}
s^2 -v(t)^2 + u(t) w(t) = 0 = s^2 + \sum_k { F_k \over t-a_k}
\end{equation}
This is the equation of a smooth hyperelliptic Riemann surface of genus
$g=N-1$~\cite{Mu84}.
We see that the $F_k$ are integrals of motion since they appear in the moduli
of the spectral curve.
Notice that $\sum_k F_k = \sum_k x_k^2$ and $H = 1/2 \,\sum_k a_k F_k$.

Since $\sum_k x_k^2$ commutes with the Hamiltonians, it generates a  
{\em canonical} transformation:
\begin{equation}
(x_k,y_k) \longrightarrow (x_k,y_k+2 \lambda x_k)
\label{invariance}
\end{equation}
which is a symmetry of the system.
The Hamiltonian reduction with respect to this symmetry is the constrained
system on the sphere.  In the Lax setup, this symmetry corresponds to
conjugating $L(t)$ by the constant matrix $$\pmatrix{1&0\cr2\lambda&1}$$  Note
that this conjugation fixes $L_0$. The properly reduced system has a phase
space of dimension $2(N-1)=2g$.

According to the standard Sklyanin procedure~\cite{Sk85}, we introduce
separating coordinates as the roots of $L_{12}(t)=u(t)$. The positivity of the
$x_i^2$ ensures that there is exactly one root in each of the intervals
$]a_k,a_{k+1}[$ so we can set:
$$a_1 \le t_1 \le a_2 \le \cdots \le a_{N-1} \le t_{N-1} \le a_N$$
The coordinates $t_j$ in this domain form an orthogonal system which
covers bijectively the quadrant $x_i \ge 0$, as illustrated in Fig.~1:
\begin{equation}
u(t)=(\sum_i x_i^2){\prod_j(t-t_j) \over \prod_k(t-a_k)} 
\quad\Longrightarrow\quad x_k^2=(\sum_i x_i^2)
{\prod_j(a_k-t_j)\over \prod_{l\neq k}(a_k-a_l)}
\label{inversion}
\end{equation}
When $t\to a_k$ we have $x_k \simeq \sqrt{t-a_k}$ so that
other quadrants are explored by analytic continuation around the branch point
at $t=a_k$.
\begin{figure}[tp]
\includegraphics[width=12cm]{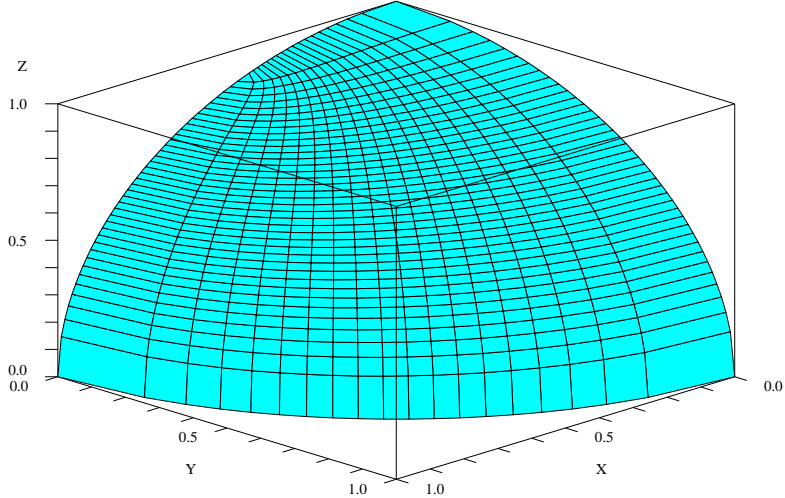}
\label{grid}\caption{A uniform grid in the separating coordinates for $N=3$.}
\end{figure}

\section{The Schr\"odinger equation.}

We shall seek wave functions which factorize in the separating variables
$t_k$:
\begin{equation}
\Psi(t_1,\cdots,t_{N-1})=\Psi_1(t_1)\,\Psi_2(t_2)\cdots\Psi_{N-1}(t_{N-1})
\label{formesep}
\end{equation}
the Hilbert space being the space of functions of variables $t_i$ with the
measure:
$${ \prod_{i>j} (t_i-t_j) \over \sqrt{\prod_{i,k} |t_i-a_k|} }\prod_i dt_i $$

It has been shown in~\cite{BaTa92} that in order to simultaneously solve the
eigenvalue equations $F_k \Psi = f_k \Psi$ all the $\Psi_k$ satisfy the same
one--dimensional Schr\"odinger equation:
\begin{equation}
\left[{d^2\over dt^2}+{1\over 2}\sum_k{1\over t-a_k}\;{d\over dt}
-{1\over 4\hbar^2}\sum_k{f_k\over t-a_k}\right]\,\Psi(t)=0
\label{lame}
\end{equation}
This is a linear differential equation with singularities at the $a_k$ and at
infinity. More precisely at each $a_k$ there is a regular
singularity~\cite{WW}, with exponents 0 and 1/2, while the singularity at
infinity is irregular. Hence this equation appears as a generalization of the
hypergeometric equation, however unlike the case of three regular
singularities there doesn't appear to exist integral formulae for the
solution, precluding to perform explicit analytic continuation around the
branch points at $a_k$.  

There exists, fortunately, a workaround. Remark
that all Hamiltonians $F_k$  are invariant under the $N$  parity
transformations $(x_j,y_j) \to (-x_j,-y_j)$ so that one can {\em
simultaneously} diagonalize both the Hamiltonians $F_k$ and the parities.
That is, we can require that the wave function is even or odd in $x_j$
for each $j$, or equivalently that is is a function of the $x_j^2$
multiplied by a monomial $\prod x_j^{n_j}$ with $n_j\in \{0,1\}$.
Let us now recall that if $t_0$ is a regular singularity with exponents
$\alpha$, $\beta$, there exists convergent developments of the solution
of the form 
$$(t-t_0)^m (p_0+p_1(t-t_0)+p_2(t-t_0)^2+\cdots)$$
where $m$ is $\alpha$ or $\beta$. The exponents being $0,\, 1/2$ at each
$a_k$, we see that these solutions are respectively even or odd under $x_k \to
- x_k$, since $x_k \simeq \sqrt{t-a_k}$ for $t \to a_k$.

Moreover let us look at the parity properties of the wave function $\Psi$ in
eq.(\ref{formesep}) with respect to $x_k$. There are two ways to approach
$x_k=0$:
$$ t_k \to a_k ~{\rm or}~ t_{k-1}\to a_k \quad
a_{k-1}<t_{k-1}<a_k<t_k<a_{k+1}$$
Requiring that the product $\Psi_{k-1}(t_{k-1}) \Psi_k(t_k)$ leads to a definite
parity state in $x_k$ amounts to saying that $\Psi_{k-1}(t)$ and $\Psi_k(t)$
share the same exponent at $a_k$, hence both share the same analytic expansion
in the interval $]a_{k-1},a_{k+1}[$.  By induction, we arrive at the stronger
statement that in order to have definite parities with respect to the $x_k$,
all functions $\Psi_k$ in eq.(\ref{formesep}) are in fact one and the same
solution $\Psi(t)$ of the differential equation~(\ref{lame}) on the whole
interval $]a_1,a_N[$, but constrained to have definite exponents at the points
$a_k$, so as to produce the above monomial $\prod x_j^{n_j}$. This is the
constraint which quantizes the $F_k$ in eq.(\ref{lame}). First it is clear
that a wave function $\Psi$ so obtained on the quadrant $x_k \ge 0$ extends by
parity to a univalued wave function on the sphere satisfying the
Schr\"odinger equation. Second, start from a solution with definite exponent
at $a_1$. In general its continuation to $a_2$ will be a superposition of
the two pure exponent solutions. Requiring that there is no mixing imposes
one condition on the $f_k$. Similarly at $a_3,\ldots,a_N$ so that we get
$N-1$ conditions on the $N-1$ independent $f_k$, the quantization conditions.

It remains to express these conditions in a form suitable for numerical
analysis. We can always reduce the problem to the case where the solution is
even under all $x_k \to -x_k$. Indeed if we want an odd solution, we have only
to write $\Psi(t)=\sqrt{t-a_k} \Phi(t)$ and obtain the differential equation
satisfied by $\Phi$, which is very similar to eq.(\ref{lame}).  Then we require
that $\Phi$ be even. Hence we need to express that the solution $\Psi(t)$
has a regular power series development at each point $a_k$. Inserting
$\Psi(t)=p_0+p_1(t-t_0)+p_2(t-t_0)^2+\cdots$ in eq.(\ref{lame}) and requiring
that the polar terms cancel we get the equation $p_1=p_0 f_k/(2\hbar^2)$,
that is:
\begin{equation}
\Psi'(a_k)=\Psi(a_k) f_k/(2\hbar^2)
\label{lamebc}
\end{equation}
Of course these equations and
eq.(\ref{lame}) are linear in $\Psi$ so a definite solution for $\Psi$ is
obtained when requiring for example a normalization condition such as 
$\Psi(a_1)=1$.

\section{Numerical techniques.}

The problem is now set in a form which happens to be directly tractable 
by a known numerical analysis package called COLNEW\footnote{This is work
of Ascher, Christiansen, Russell, and Bader, which can be found at
http://www.netlib.org/ode/}. This package
allows to solve so--called mixed order boundary values problems:
suppose we have a set of differential equations of various orders
$u_i^{(m_i)}=f_i(u)$ where $u$ is a vector of all functions $u_i$
and all their derivatives up to order $m_i -1$ for $u_i$, where $f$
may be linear or non linear. The general solution depends 
on $\sum m_i$ constants. The package allows to impose $\sum m_i$
``boundary conditions'' and determines the solution satisfying all of them.
These boundary conditions have to be of the form $g_j(u)(\zeta_j)=0$
where $g_j$ are expressions of the same vector $u$ as above, and the points
$\zeta_j$ are $\sum m_i$ points arbitrarily disposed in the considered domain
or its boundary. The trick to apply this scheme to our problem is to treat
the quantities $f_k$ as functions of $t$ which happen to be constant, i.e.
satisfy $f_k'=0$. Supplementing eq.(\ref{lame}) by these $N-1$ independent
conditions we get a mixed order non linear differential system with total
order $2+(N-1).1=N+1$. The boundary conditions we impose are the $N$
above equations $\Psi'(a_k)=\Psi(a_k) f_k/(2\hbar^2)$ at each $a_k$ plus
a normalization such as $\Psi(a_1)=1$ which completely fixes things.

Finally to appreciate how the numerical package can deal with the
singularities at the $a_k$ appearing in eq.(\ref{lame}), it is in
order to say a few words about it. The computation starts by cutting
the considered interval into a number of subintervals. One is free to require
that the $a_k$ are included in the subdivision. In each interval, the solution
is approximated by some polynomial. The coefficients of these polynomials are
fixed by a set of equations: first one requires that the differential equation
is exactly satisfied at a number of so--called ``collocation points'' inside the
intervals.  These points are chosen at Gaussian positions as in numerical
integration, in order to improve convergence. Note that, these points being
interior while the $a_k$ are boundary points, we are always  away from the
poles in $1/(t-a_k)$ when computing these equations. Second the package
imposes that the solutions from interval to interval fit in a smooth way, and
that the boundary conditions are obeyed. These conditions don't suffer from
divergent factors at $a_k$ and moreover imply in a finite way that the
differential equation is indeed obeyed at the $a_k$ and that the solution has
the appropriate parity property.

Things are set up so that one gets a non linear system for the coefficients,
with as many equations as unknowns. This system is then solved by
Newton--Raphson method, an error is evaluated, and a more refined solution is
computed until a defined tolerance is obtained on all the equations. In
practice we have found that the package works remarkably well for our problem,
and is able to compute a fair number of solutions in a couple of seconds on a
modern machine. Moreover its use is enhanced by the existence of a wrapper in
Scilab\footnote{http://scilabsoft.inria.fr/}
which allows to program these computations in a convenient way.

The numerical results below will be restricted to the case $N=3$.  There would
not be any difficulty to study similarly higher dimensional problems.  To
reduce the parameter space to explore, we first normalize coefficients.
Translating all the $a_k$ by the same quantity does not change the $F_k$ and
adds a constant to the energy.  We therefore set $a_1=0$.  The
equation~(\ref{lame}) is also invariant by a common rescaling of $t$ and the
$a_k$ by $\lambda$ and the rescaling of the $f_k$ by $1/\lambda$.  The $\hbar$
factor can also be absorbed in the $f_k$ so that we are reduced to two
parameters.  We choose $a_2=1$ and have a parameter $y=a_3/a_2$ characterizing
the asymmetry of the model.  Now $\sum_k f_k =v$, where
\begin{equation}
 v={a_2 r^2 \over 4 \hbar^2}, \quad r^2=\sum_k x_k^2
 \label{defv}
\end{equation}
This single parameter can either be seen as characterizing the strength of the
potential energy compared to the kinetic energy, or the approach to the
semiclassical regime, which are therefore equivalent in this case.
Finally, let us write the equations to be solved:
\begin{eqnarray}
&&\hskip -1cm \biggl( {d^2\over dt^2} +{1\over 2}\Bigl({1\over t} + {1\over t-1}
+ {1\over t-y} \Bigr)  {d\over dt} -\nonumber\\
&&\hskip 1cm\Bigl({v-f_2-f_3\over t} + {f_2\over t-1}
+ {f_3\over t-y} \Bigr) \biggr) \Psi(t) = 0 \label{wangerin}\\
&&\hskip -1cm \biggl( {d^2\over dt^2} +{1\over 2}\Bigl({3\over t} + {1\over t-1}
+ {1\over t-y} \Bigr)  {d\over dt} -\nonumber\\
&&\hskip -.5cm \Bigl({v-f_2-f_3+1/4+1/(4y)\over t} +
{f_2-1/4\over t-1} + {f_3-1/(4y)\over t-y} \Bigr) \biggr) \Psi(t) = 0 
\end{eqnarray}
The second equation is the one we obtain when setting
$\Psi \to \sqrt t \Psi$.  There exist six similar equations describing, with
the two displayed, the eight possible combinations of parities of the solution.
Equation~(\ref{wangerin}) is known as Wangerin's equation, but its properties
have not been explored. When $v=0$ the singularity at infinity becomes
regular, and the equation reduces to Lam\'e's equation. A great deal of work
on this equation is summarized in~\cite{WW}, and is relevant for our study
as we see below.

\section{Numerical results.}

\begin{figure}
\hskip -1cm \includegraphics[width=14cm]{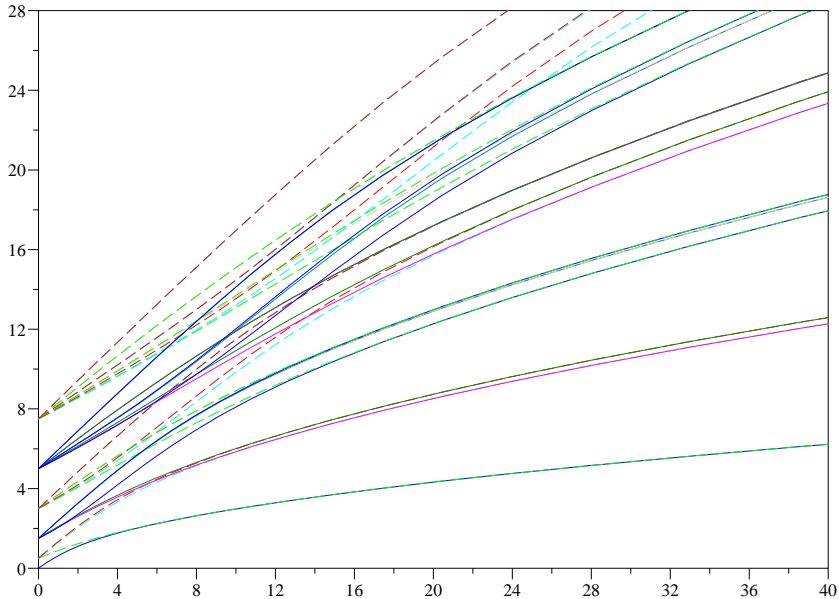}
\caption{Energy as a function of $v$. Case $y=1.1$.}
\end{figure}

A single spectrum is not really instructive. We shall therefore show the
evolution of the 36 lowest energy levels when the parameter $v$ grows.
The number of different energy levels displayed is limited for the readability
of the plots.  States are characterized by their sets of parities and by the
number of zeros of the wave function in each of the two intervals $]0,1[$
and $]1,y[$.  The principal difficulty is to be able to choose a particular
solution with a given number of zeros. The COLNEW package allows for a powerful
mean to achieve this: an initial guess of the solution can be provided.
However good guesses are not easy to find except for $v=0$. In this case, the
theory of the Lam\'e spheroidal harmonics teaches us that the solutions which
are analytic at the $a_k$ are polynomials.  The zeros of these polynomials
satisfy simple equations and it is therefore not difficult to give numerical
approximations of the solutions.  To reach higher values of $v$, we have used
the continuation method, which consists in taking the solution for the old
value of $v$ as the initial guess for the next value of $v$.  If the
difference between successive values of $v$ is sufficiently small, the new
solution has the same quantum numbers\footnote{The standard COLNEW wrapper in
Scilab is however unsuitable to perform this continuation and we had to modify
it.}. Below we have plotted $E/2 = f_2+yf_3$ but we could have
plotted individually each $f_k$ as a function of $v$.

To have an idea of the effect of the asymmetry parameter $y$, we plot the
energies for three values of this parameter.  The line styles correspond to
the parity properties of the states. The heavy solid lines correspond
to the states which are completely even, the other lines are solid or dashed
according to the total parity of the states.  In Fig.~2, $y$ is equal to
1.1, meaning that the two coordinates $x_2$ and $x_3$ are nearly equivalent.
In fact, when $a_k=a_{k+1}$ there is a singular term in $F_k$ and $F_{k+1}$,
but we can change for their sum and difference, with the sum being regular and
the difference proportional to $J_{k,k+1}^2$.  In this case of enhanced
symmetry, the algebro--geometric solution of the classical model becomes
singular, see~\cite{BBT}, but the quantum solution has no particular trouble.
Note however that a degeneracy of energy levels is apparent on the figure.
\begin{figure}
\hskip -1cm \includegraphics[width=14cm]{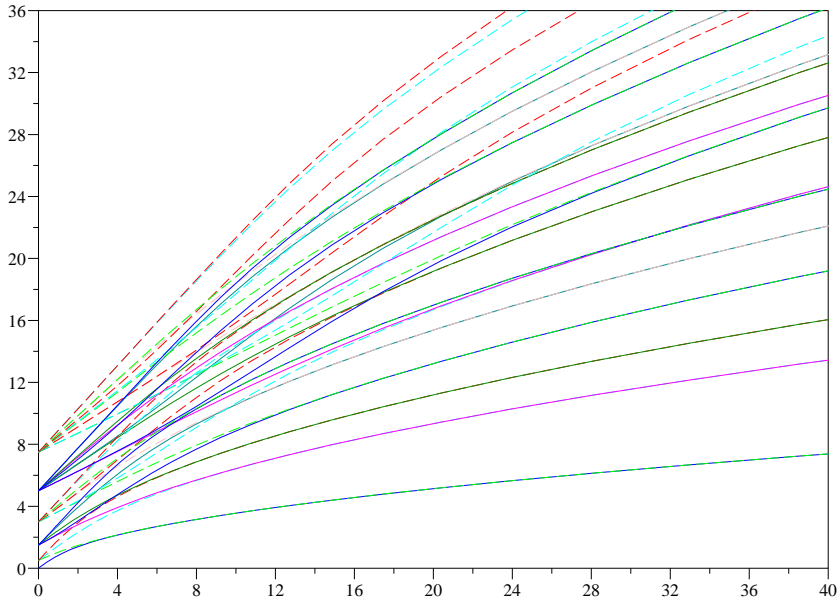}
\caption{Energy as a function of $v$. Case $y=2$.}
\end{figure}
The case $y=2$ in Fig.~3 is representative of the case where the oscillator strengths
are equally distributed, while the case $y=5$ in Fig.~4 illustrates the case where
one of them is bigger. A remarkable feature of these plots is that the
levels converge at $v=0$ to integer or half integer values. This is easily
understood:  $v=0$ is equivalent to setting the potential to zero, hence the
problem has spherical symmetry, so the energy is equal to a factor
times the eigenvalue of the $L^2$ operator, $j(j+1)$ with a $2j+1$
degeneracy. This is easily observed on the curves. An other feature of these
plots is that pair of levels converge for large $v$. This can be understood
from the potential barrier at the $x_1=0$ plane, so that the probability is
concentrated around the opposite poles.  In this limit, there are therefore
two equivalent states, just separated by the exponentially small tunneling
probability.  When $v$ becomes even greater, the states occupy but a small
patch around the poles and the effect of the curvature of the sphere becomes
small.  The problem reduces to the one of two independent harmonic
oscillators, which have energies proportional to $\sqrt v$.  Remark however
that with growing quantum numbers, the level crossings between the two regimes
$v$ small and $v$ large occur at larger values of $v$, so that for a given
value of $v$, the eigenfunctions of sufficient energies are
perturbations of the spheroidal harmonics.

\begin{figure}
\hskip -1cm \includegraphics[width=14cm]{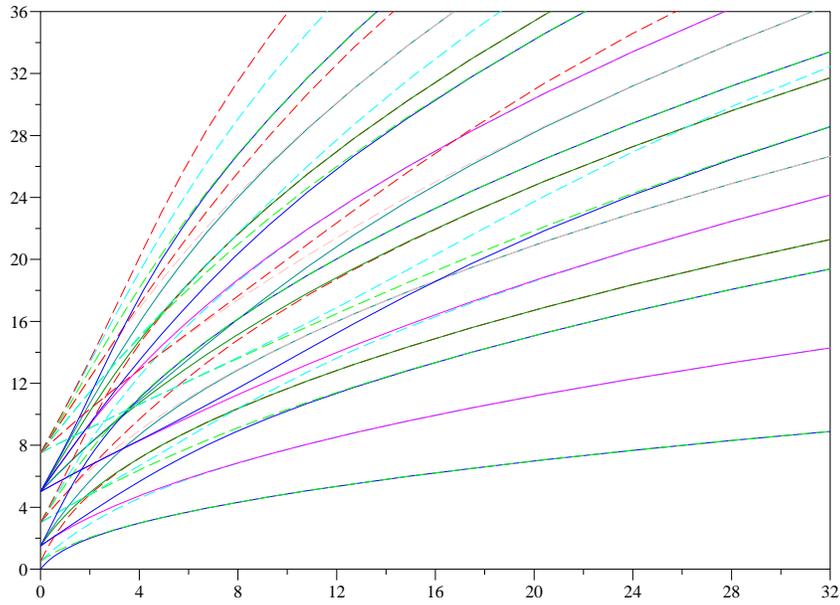}
\caption{Energy as a function of $v$. Case $y=5$.}
\end{figure}

\begin{figure}
\begin{center}
\includegraphics[width=14cm]{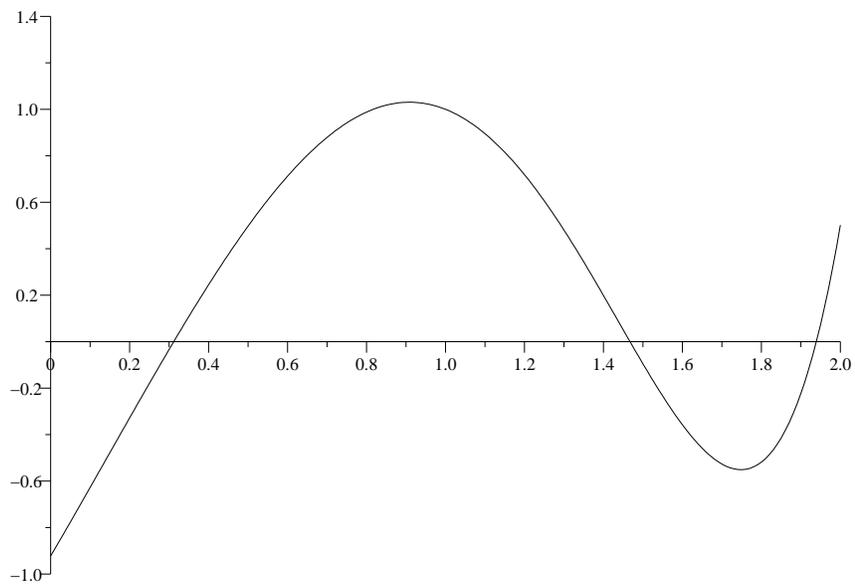}
\end{center}
\label{sol}
\caption{Eigenfunction. Case $y=2$ and $v=8$.}
\end{figure}

Finally the solution of eq.~(\ref{lame}) for an excited state with $v=2$ and
$y=2$ is plotted in figure~(\ref{sol}).  Precisely for this solution, which
has one zero in $]0,1[$ and two zeroes in $]1,y[$ we have $f_2=-0.338$,
$f_3=9.913$ hence the "energy" is $f_2+yf_3=19.49$.  Recall that the wave
function is $\Psi(t_1)\Psi(t_2)$, with $0\le t_1\le 1 \le t_2 \le y$. Note
that the locus of zeroes of the wave function is formed of intersecting lines
due to the separation of variables.  This is clearly non generic and would be
destroyed by any perturbation.  The existence of such singular node curves in
the eigenstates of a quantum Hamiltonian is evidence for integrability.

\section{Conclusion.}
An efficient calculation of the spectrum of the quantum integrable Neumann
model has been displayed.  A number of degeneracies are evident from the
above figures.  It will be the purpose of a separate work to explicit these
phenomena and relate them with the semiclassical analysis of the equation.
This makes contact with the solution of the classical case in terms
of hyperelliptic functions. The interesting features are the transitions
between different semiclassical quantization conditions.

\end{document}